\documentclass[sn-mathphys,Numbered]{sn-jnl}

\usepackage{graphicx}%
\usepackage{multirow}%
\usepackage{amsmath,amssymb,amsfonts}%
\usepackage{amsthm}%
\usepackage{mathrsfs}%
\usepackage[title]{appendix}%
\usepackage{xcolor}%
\usepackage{textcomp}%
\usepackage{manyfoot}%
\usepackage{booktabs}%
\usepackage{algorithm}%
\usepackage{algorithmicx}%
\usepackage{algpseudocode}%
\usepackage{listings}%
\usepackage{natbib}

\usepackage{float}
\usepackage{comment}

\theoremstyle{thmstyleone}%
\theoremstyle{thmstyletwo}%

\theoremstyle{thmstylethree}%

\begin{document}

\title[Article Title]{Broadband parametric amplification in DARTWARS}


\author*[1,2,3]{\fnm{M.} \sur{Faverzani}}\email{marco.faverzani@unimib.it}
\equalcont{These authors contributed equally to this work.}
\author[1,2,3]{\fnm{P.} \sur{Campana}}
\equalcont{These authors contributed equally to this work.}
\author[1,2,3]{\fnm{R.} \sur{Carobene}}
\equalcont{These authors contributed equally to this work.}
\author[1,2,3]{\fnm{M.} \sur{Gobbo}}
\equalcont{These authors contributed equally to this work.}
\author[4,5]{\fnm{F.} \sur{Ahrens}}
\author[6]{\fnm{G.} \sur{Avallone}}
\author[6,7]{\fnm{C.} \sur{Barone}}
\author[1,2,3]{\fnm{M.} \sur{Borghesi}}
\author[1,2]{\fnm{S.} \sur{Capelli}}
\author[6,7]{\fnm{G.} \sur{Carapella}}
\author[8,9]{\fnm{A. P.} \sur{Caricato}}
\author[10]{\fnm{L.} \sur{Callegaro}}
\author[11,12]{\fnm{I.} \sur{Carusotto}}
\author[1]{\fnm{A.} \sur{Celotto}}
\author[4,5]{\fnm{A.} \sur{Cian}}
\author[13]{\fnm{A.} \sur{D'Elia}}
\author[13]{\fnm{D.} \sur{Di Gioacchino}}
\author[10,5]{\fnm{E.} \sur{Enrico}}
\author[4,14,5]{\fnm{P.} \sur{Falferi}}
\author[10]{\fnm{L.} \sur{Fasolo}}
\author[2]{\fnm{E.} \sur{Ferri}}
\author[15,7]{\fnm{G.} \sur{Filatrella}}
\author[13]{\fnm{C.} \sur{Gatti}}
\author[4,5]{\fnm{D.} \sur{Giubertoni}}
\author[6,7]{\fnm{V.} \sur{Granata}}
\author[6,7]{\fnm{C.} \sur{Guarcello}}
\author[1,4,5]{\fnm{A.} \sur{Irace}}
\author[1,2,3]{\fnm{D.} \sur{Labranca}}
\author[8,9]{\fnm{A.} \sur{Leo}}
\author[13]{\fnm{C.} \sur{Ligi}}
\author[13]{\fnm{G.} \sur{Maccarrone}}
\author[4,5]{\fnm{F.} \sur{Mantegazzini}}
\author[4,5]{\fnm{B.} \sur{Margesin}}
\author[8,9]{\fnm{G.} \sur{Maruccio}}
\author[12,5]{\fnm{R.} \sur{Mezzena}}
\author[8,9]{\fnm{A. G.} \sur{Monteduro}}
\author[1,2,3]{\fnm{R} \sur{Moretti}}
\author[1,2,3]{\fnm{A.} \sur{Nucciotti}}
\author[10,5]{\fnm{L.} \sur{Oberto}}
\author[1,2,3]{\fnm{L.} \sur{Origo}}
\author[6,7]{\fnm{S.} \sur{Pagano}}
\author[13]{\fnm{A. S.} \sur{Piedjou Komnang}}
\author[13]{\fnm{L.} \sur{Piersanti}}
\author[13]{\fnm{A.} \sur{Rettaroli}}
\author[8,9]{\fnm{S.} \sur{Rizzato}}
\author[13]{\fnm{S.} \sur{Tocci}}
\author[14,4,5]{\fnm{A.} \sur{Vinante}}
\author[1,2,3]{\fnm{M.} \sur{Zannoni}}
\author[1,2,3]{\fnm{A.} \sur{Giachero}}


\affil[1]{\orgdiv{Physics Department}, \orgname{University of Milano - Bicocca}, \orgaddress{\city{Milano}, \country{Italy}}}
\affil[2]{\orgname{INFN Sezione di Milano - Bicocca}, \orgaddress{\city{Milano}, \country{Italy}}}
\affil[3]{\orgname{Bicocca Quantum Technologies (BiQuTe) Centre}, \orgaddress{\city{Milano}, \country{Italy}}}
\affil[4]{\orgname{Fondazione Bruno Kessler}, \orgaddress{\city{Trento}, \country{Italy}}}
\affil[5]{\orgname{INFN - TIFPA}, \orgaddress{\city{Trento}, \country{Italy}}}
\affil[6]{\orgdiv{Physics Department} \orgname{University of Salerno}, \orgaddress{\city{Salerno}, \country{Italy}}}
\affil[7]{\orgname{INFN Sezione di Napoli}, \orgaddress{\city{Napoli}, \country{Italy}}}
\affil[8]{\orgdiv{Physics Department} \orgname{University of Salento}, \orgaddress{\city{Lecce}, \country{Italy}}}
\affil[9]{\orgname{INFN Sezione di Lecce}, \orgaddress{\city{Lecce}, \country{Italy}}}
\affil[10]{\orgname{INRiM}, \orgaddress{\city{Torino}, \country{Italy}}}
\affil[11]{\orgname{INO-CNR BEC Center}, \orgaddress{\city{Trento}, \country{Italy}}}
\affil[12]{\orgdiv{Physics Department} \orgname{University of Trento}, \orgaddress{\city{Trento}, \country{Italy}}}
\affil[13]{\orgname{INFN Laboratori Nazionali di Frascati}, \orgaddress{\city{Frascati}, \country{Italy}}}
\affil[14]{\orgname{IFN-CNR}, \orgaddress{\city{Trento}, \country{Italy}}}
\affil[15]{\orgdiv{Department of Science and Technology} \orgname{University of Sannio}, \orgaddress{\city{Benevento}, \country{Italy}}}

\abstract{Superconducting parametric amplifiers offer the capability to amplify feeble signals with extremely low levels of added noise, potentially reaching quantum-limited amplification. This characteristic makes them essential components in the realm of high-fidelity quantum computing and serves to propel advancements in the field of quantum sensing. In particular, Traveling-Wave Parametric Amplifiers (TWPAs) may be especially suitable for practical applications due to their multi-Gigahertz amplification bandwidth, a feature lacking in Josephson Parametric Amplifiers (JPAs), despite the latter being a more established technology. This paper presents recent developments of the DARTWARS (Detector Array Readout with Traveling Wave AmplifieRS) project, focusing on the latest prototypes of Kinetic Inductance TWPAs (KITWPAs).
The project aims to develop a KITWPA capable of achieving $20\,$ dB of amplification. To enhance the production yield, the first prototypes were fabricated with half the length and expected gain of the final device.
In this paper, we present the results of the characterization of one of the half-length prototypes. The measurements revealed an average amplification of approximately $9\,$dB across a $2\,$GHz bandwidth for a KITWPA spanning $17\,$mm in length.

}




\maketitle

\section{Introduction}\label{sec1}

The rapid evolution of quantum computing demands for amplifiers featuring wide bandwidth, large dynamic range, and quantum-limited added noise. Devices with these characteristics would also offer significant advantages to quantum sensing for the readout of low-temperature detectors, ultimately enhancing the performance of next-generation experiments.

Superconducting parametric amplifiers exploit a non-linear circuit to amplify signals. Among these devices, Josephson Parametric Amplifiers (JPAs)~\cite{JPA} stand out as the best-established technology. While JPAs have demonstrated their ability to achieve quantum-limited noise performances, they are constrained by other limitations, primarily associated with their relatively narrow bandwidth and limited dynamic range of amplification.

Another class of devices comprises Traveling Wave Parametric Amplifiers (TWPAs), which have the potential to expand the bandwidth and dynamic range to values close to the ones of existing standard commercial amplifiers. A TWPA features a non-linear transmission line that is either embedded with Josephson junctions (TWJPA)~\cite{Macklin2015} or made with high-kinetic inductance materials (KITWPA)~\cite{HoEom2012}. 
Both of these approaches are being pursued~\cite{Borghesi2022,Rettaroli2022,Pagano,Granata,Guarcello} by the DARTWARS (Detector Array Readout with Traveling Wave AmplifieRS) project, which aims to reach gain, noise and saturation power of around $20\,$dB, $600\,$mK, and $-50\,$dBm respectively.

\begin{figure}[t!]
    \centering
    \begin{minipage}[t]{0.48\textwidth}
        \centering
        \includegraphics[width=0.9\linewidth]{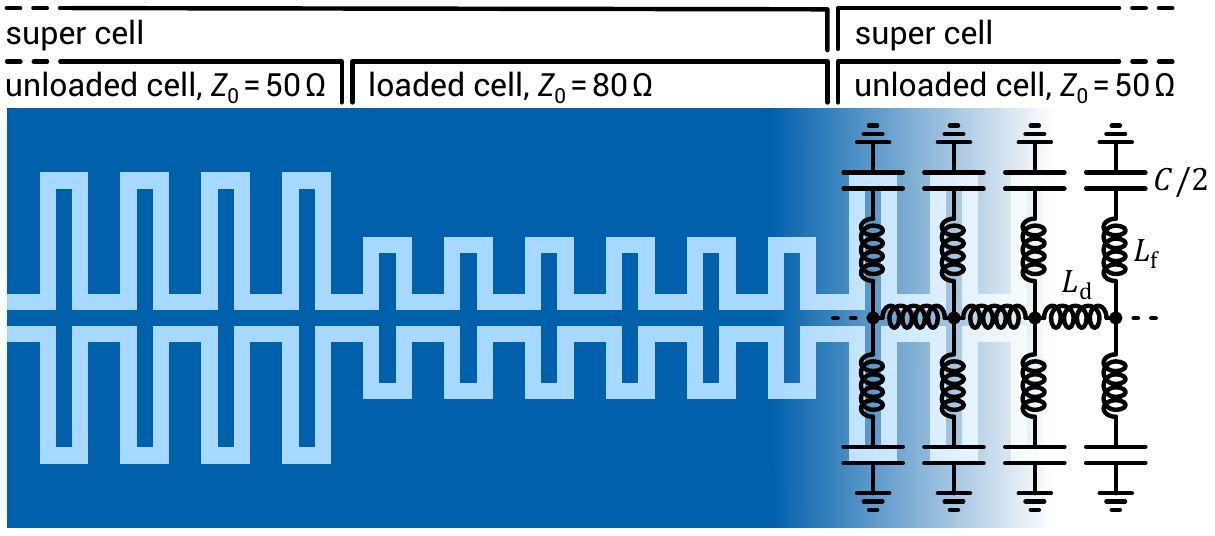}
        \caption{The diagram (not to scale) of the lumped element artificial transmission line. The dark blue areas indicate the high-resistivity superconductor, whereas the light blue areas stand for the substrate.}
        \label{fig:fishbone}
    \end{minipage}\hfill
    \begin{minipage}[t]{0.48\textwidth}
        \centering
        \includegraphics[width=0.9\linewidth]{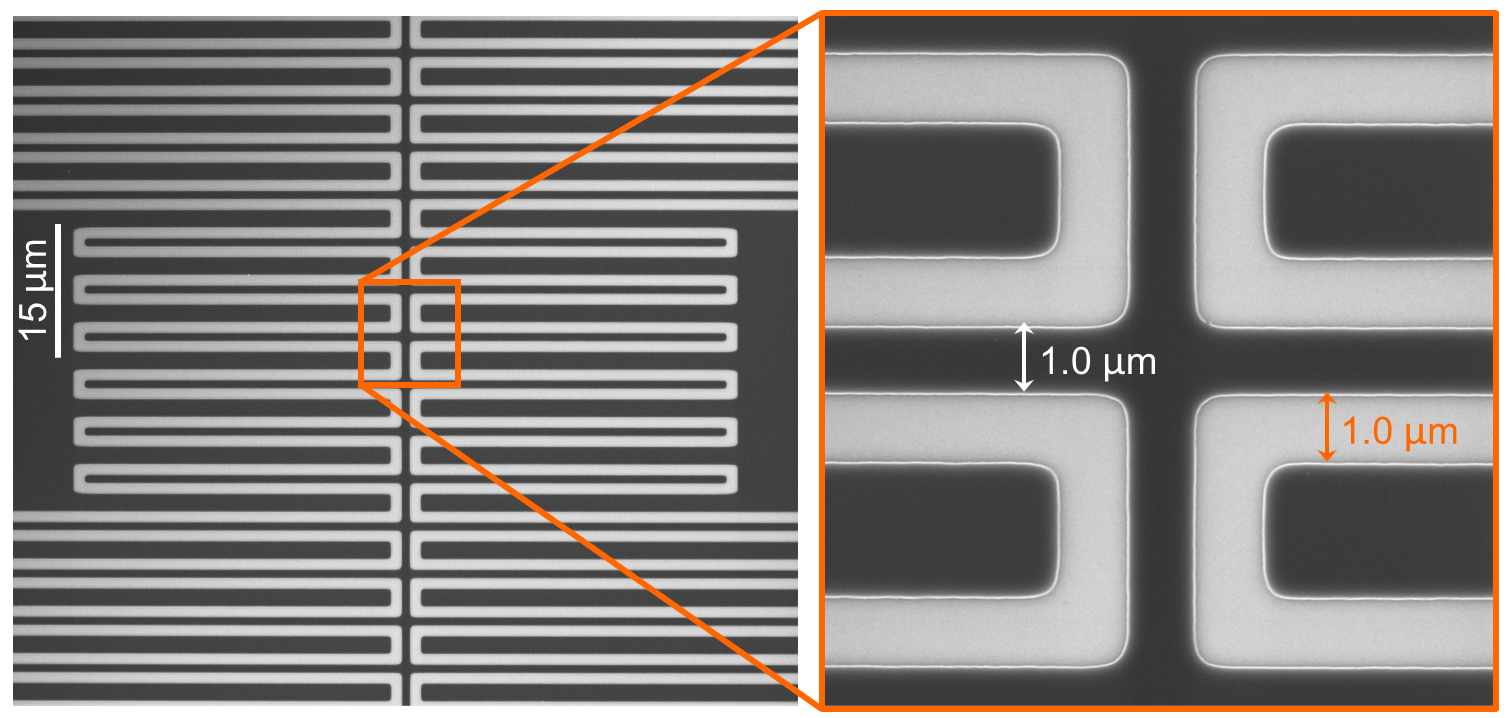}
        \caption{Scanning electron microscope images of the patterned artificial transmission line. Dark areas represent the NbTiN film, whereas light areas are the Si substrate.}
        \label{fig:SEM}
    \end{minipage}
\end{figure}

This paper focuses on the latest developments of DARTWARS, providing a more comprehensive characterization of the KITWPA prototypes introduced in~\cite{Mantegazzini_2023}, which demonstrated significant amplification in preliminary tests.
A KITWPA consists of a weakly dispersive phase-matched transmission line, controlled by dispersion engineering, which is required to create exponential amplification~\cite{Chaudhuri2017}.
The prototype characterized in this paper is an artificial transmission line organized in multiple cells, each one composed of a stub-loaded coplanar waveguide (CPW) with an inductance ($L$) dependent on the kinetic inductance of the material and two interdigitated capacitors (IDC) to establish capacitance ($C$) towards the ground.
Dispersion engineering was achieved by organizing the transmission line into $N_{sc}=523$ super-cells, each comprising $60$ \textit{unloaded cells} with a characteristic impedance $Z_0=\sqrt{{L}/{C}}=50\,\Omega$, and $6$ \textit{loaded cells} with $Z_0=80\,\Omega$, for a total of $L_{sc}=66$ cells in each supercell (fig. \ref{fig:fishbone}).

The $17\,$mm transmission line length corresponds to an expected amplification of approximately $10\,$dB. The design of the KITWPAs was optimized to produce these devices with a bandwidth centered at $4-5\,$GHz. The next fabrication will produce devices with a higher ratio of loaded to unloaded cells, shifting the bandwidth center up to $6\,$GHz.
The KITWPAs were produced by depositing a thin film of NbTiN on a high-resistivity silicon wafer (fig. \ref{fig:SEM}). Further details on the fabrication process are reported in~\cite{Mantegazzini_2023}.

\begin{figure}
    \centering
    \includegraphics[width=0.9\linewidth]{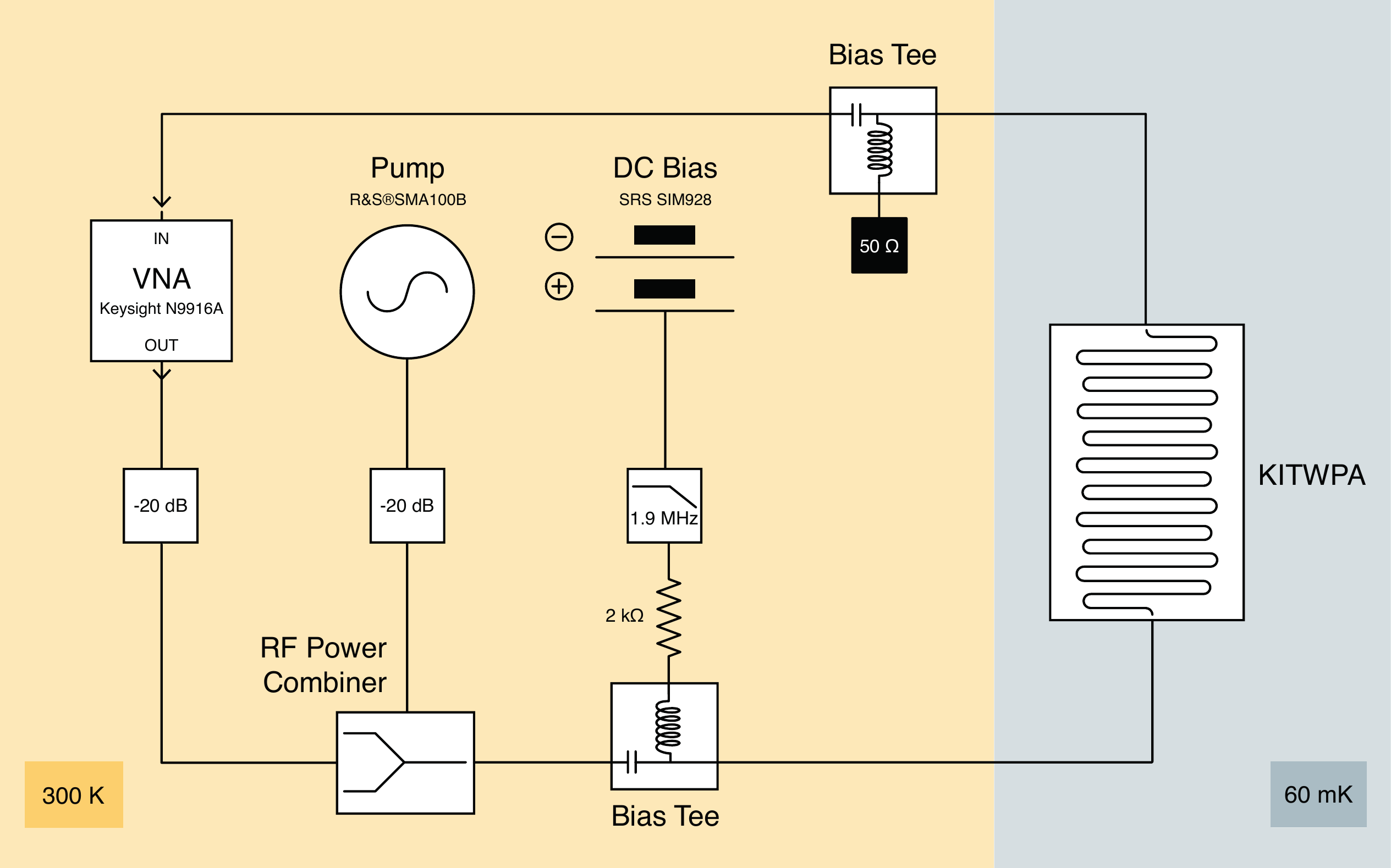}
    \caption{A schematic diagram of the experimental setup. At room temperature, the VNA and pump rf signals are combined using a power combiner. A dc current produced by an isolated voltage source is then added through a bias tee. The resulting rf/dc combined signal is routed inside the cryostat to the KITWPA located on the mixing chamber at $60\,$mK. The output signal is terminated with a $50\,\Omega$ load using a second bias tee at room temperature before the input port of the VNA.}
    \label{fig:expsetup}
\end{figure}

\section{Characterization of the KITWPA prototype}
In the present work, one of the devices with $I_c>1\,$mA and a clearly identifiable stop-band was selected for low-temperature characterization at $60\,$mK.
The experimental setup shown in fig. \ref{fig:expsetup} was used to scan the $S_{21}$ response of the TWPA while controlling its dc bias and pump signal. 
The instruments used for performing the characterization were controlled using the \texttt{Qtics} Python package~\cite{carobene_2024_10450507}. The KITWPA chip was mounted in a general-purpose copper box, equipped with CPW launchers to feed the device.

\subsection{Measurement of non-linearity parameters}
At the lowest order, the inductance of the KITWPA increases quadratically as a function of the dc bias $I$. This non-linear response can be measured by considering the phase of the $S_{21}$ parameter at a fixed frequency $f$: 

\begin{equation}\label{eq:istar}
    \frac{\theta(I)-\theta_0}{\theta_{r}} = - \frac{1}{2} \left(\frac{I}{I^*}\right)^2,
\end{equation}

where $\theta_0$ is the phase at zero bias, while $I^*$ is an intrinsic parameter of the material known as the scaling current. Finally, $\theta_r=2\pi f \Delta t_{\text{TWPA}}$ represents the phase delay over the entire  transmission line.  

The time required by a signal to travel across the device $\Delta t_{\text{TWPA}}=33\,$ns was measured with a Time-Domain-Reflectometer (TDR). This allowed to estimate the phase delay and the phase velocity $v_p = (N_\text{sc}L_\text{sc})/\Delta t_\text{TWPA}$. The latter was also cross-checked with an alternative estimate based on the relation $v_p = \lambda_{sb}\, f_{sb}$, where $\lambda_{sb} = 2L_{sc}$ and $f_{sb}=7.7\,$GHz refer respectively to the wavelength corresponding to the spatial separation between subsequent periodic loadings and the central frequency of the measured stop-band. These approaches yielded a phase-velocity of $v_p = 1016\,\text{cells/ns}$ ($1.67\cdot 10^{-3}\,c$) and $v_p = 1030\,\text{cells/ns}$ ($1.69\cdot 10^{-3}\,c$) respectively, demonstrating consistent results between the two techniques.

\begin{figure}[t!]
    \centering
    \begin{minipage}[t]{0.48\textwidth}
        \centering
        \includegraphics[width=\linewidth]{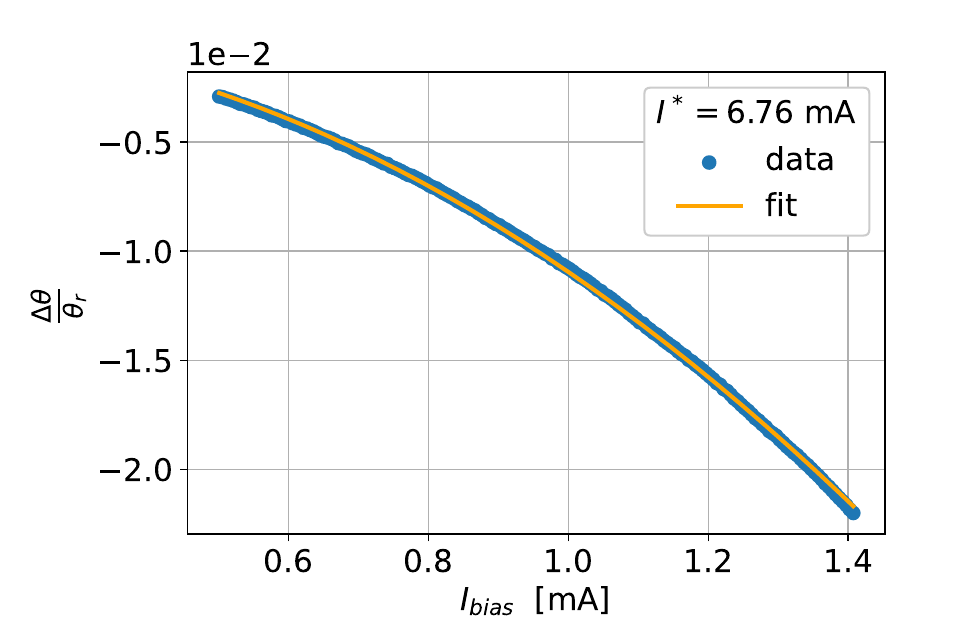}
        \caption{Measured relative phase of a signal transmitted through the TWPA with the pump off as a function of bias current. The quadratic fit allows to estimate the scaling current $I^*$.}
        \label{fig:istar}
    \end{minipage}\hfill
    \begin{minipage}[t]{0.48\textwidth}
        \centering
        \includegraphics[width=\linewidth]{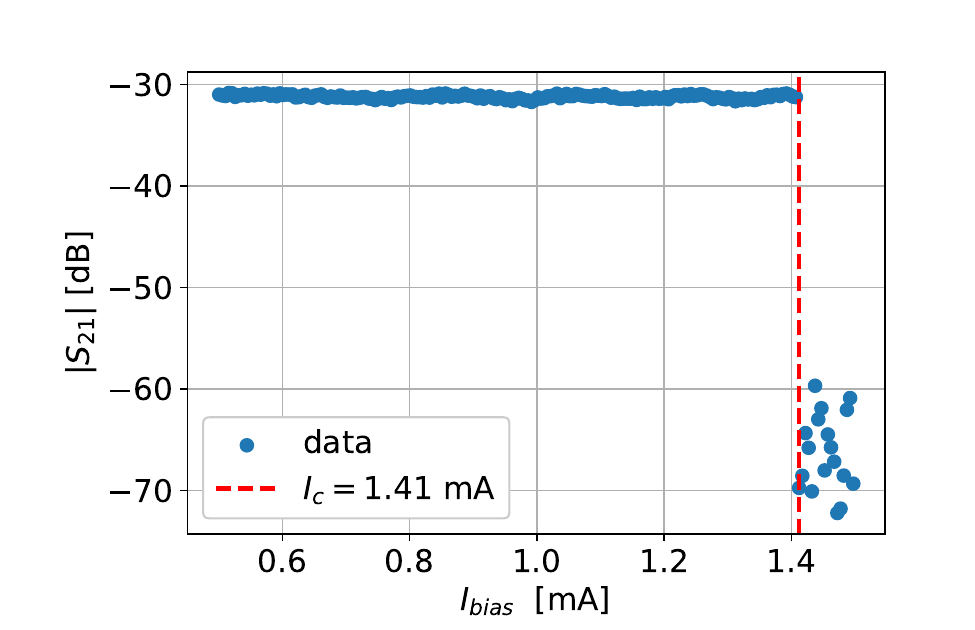}
        \caption{Measured amplitude of $S_{21}$ with the pump off as a function of bias current. The critical current $I_c$ is identified by a sharp discontinuity in the values.}
        \label{fig:icrit}
    \end{minipage}
\end{figure}

To measure the scaling current $I^*$, the data were acquired by recording the $S_{21}$ response with the VNA for various dc bias values while keeping the rf pump turned off. 
 As shown in fig. \ref{fig:istar}, fitting the measured and rescaled phase at a fixed frequency with eq. \ref{eq:istar} allows to determine $I^*$. Repeating the fit for different frequencies over the whole amplification bandwidth resulted in an average value of $I^*=(6.72 \pm 0.03)\,$mA, where the uncertainty was estimated as the standard deviation of the fit results.

The critical current $I_c$ of the device was instead measured by considering the amplitude of $S_{21}$ as a function of the dc bias. For a fixed frequency, $I_c$ can be easily identified as a significant discontinuity in the transmitted VNA signal amplitude (fig. \ref{fig:icrit}). Consistent values of $I_c$ were found when considering different frequencies across the entire amplification bandwidth, resulting in $I_c = (1.41 \pm 0.01)\,$mA.

In a preliminary characterization~\cite{Mantegazzini_2023}, another batch of prototypes was tested in liquid helium, which led to a measured scaling and critical currents of $I^*=(5.3 \pm 0.1)\,$mA and $I_c=(1.5\pm 0.2)\,$mA respectively.
Comparing these measurements with our device, the critical currents are in good agreement while the scale currents have slightly different values, but still maintain the same order of magnitude.

\subsection{Gain measurements}

\begin{figure}
    \centering
    \begin{minipage}[t]{0.48\textwidth}
        \centering
        \includegraphics[width=\linewidth]{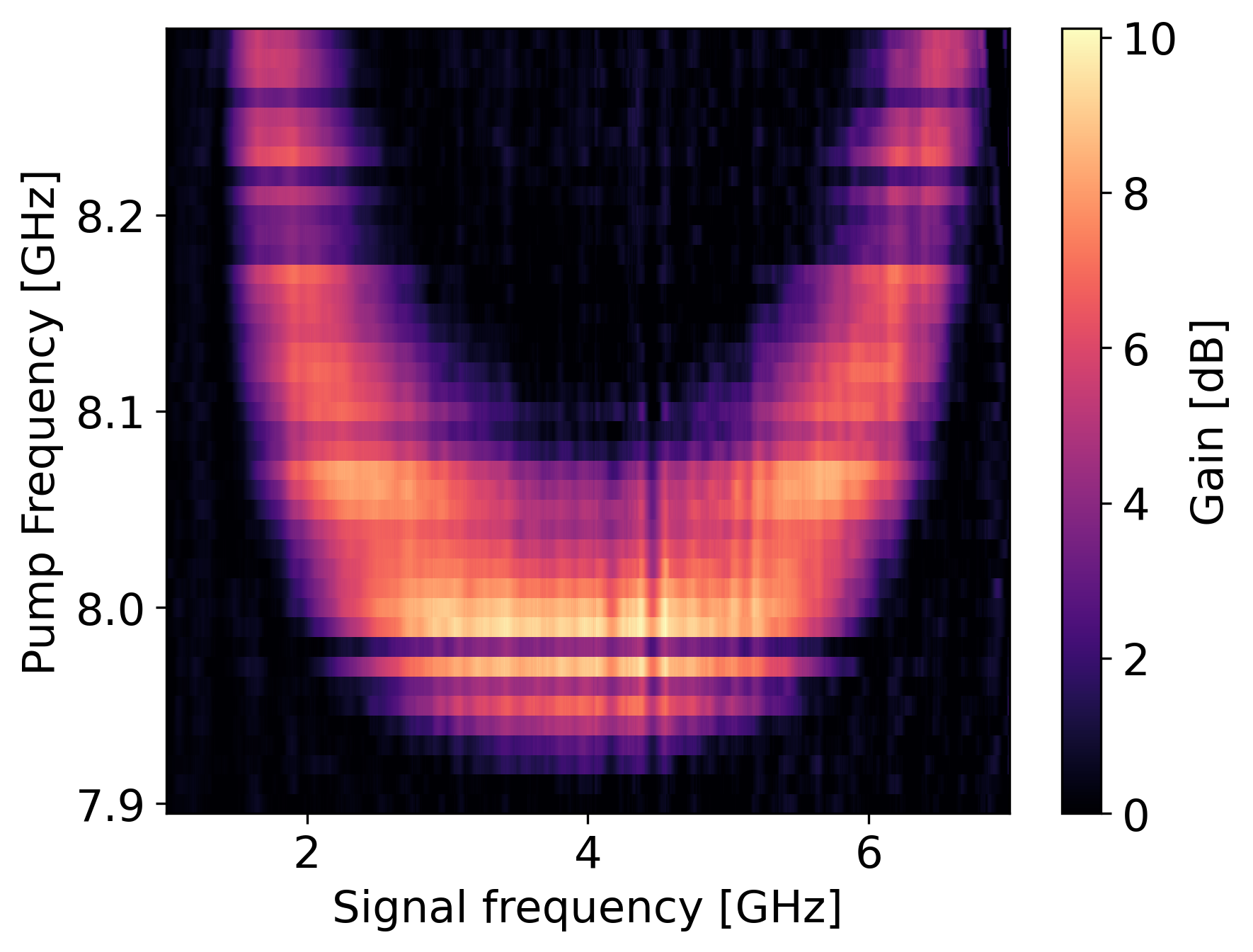}
        \caption{Measured gain as a function of VNA rf signal frequency and pump frequency. Near a pump frequency of $8.1\,$GHz, a bandwidth splitting can be observed}
        \label{fig:gain-vs-vna-frequency-vs-pump-frequency}
    \end{minipage}\hfill
    \begin{minipage}[t]{0.48\textwidth}
        \centering
        \includegraphics[width=\linewidth]{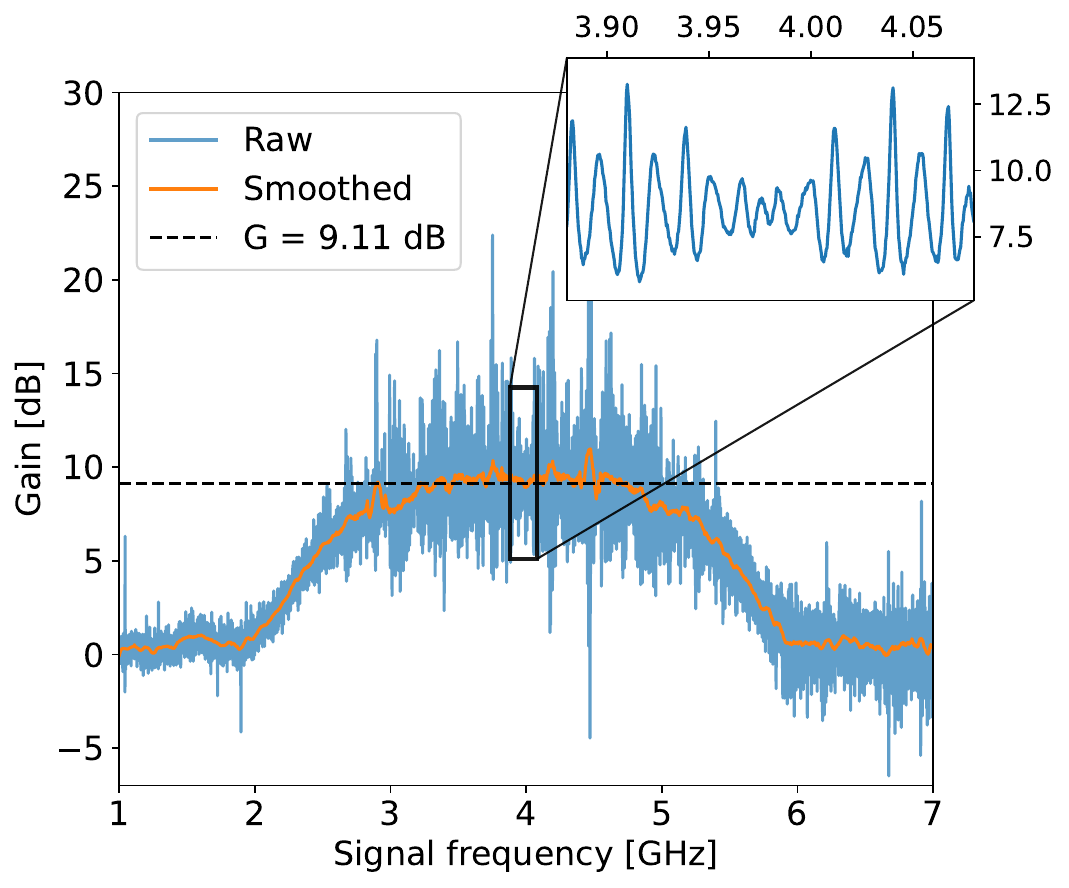}
        \caption{Maximal gain profile in a $1-7\,$GHz frequency range. The apparently noisy profile is instead produced by large ripples, as shown in the inset.}
        \label{fig:gain-vs-vna-frequency}
    \end{minipage}
\end{figure}

The best gain and corresponding bandwidth were characterized through a VNA frequency sweep in the $1-8\,$GHz range, while varying the pump rf signal within the $7.9 - 8.3\,$GHz range, which is just above the stop-band. Additional parameters, such as the dc biasing current or the pump power, were fine-tuned separately to $1.2\,$ mA and $-39.4\,$ dBm.
In fig. \ref{fig:gain-vs-vna-frequency-vs-pump-frequency}, the measured gain is plotted as a function of the VNA and the pump rf signals frequencies. As predicted by the phase-matching relations~\cite{Malnou2021}, a splitting of the bandwidth becomes apparent as the pump rf signal frequency increases.
A mean gain of $9.11\,$dB was achieved over a $3-5\,$GHz range at a pump rf signal frequency of $7.99\,$GHz (fig. \ref{fig:gain-vs-vna-frequency}).
However, the gain does not appear to be uniform due to large ripples (inset in fig. \ref{fig:gain-vs-vna-frequency}).
By performing a Fast Fourier Transform (FFT) of the transmitted signal, it can be seen that the dominant frequency corresponds to a wavelength approximately twice the length of the device.
This explains the ripples as resulting from impedance mismatches introduced by the line adaptation and connections between the KITWPA and the packaging, which lead to the creation of standing waves across the device.

\begin{figure}[t]
\centering
    \includegraphics[width=0.5\textwidth]{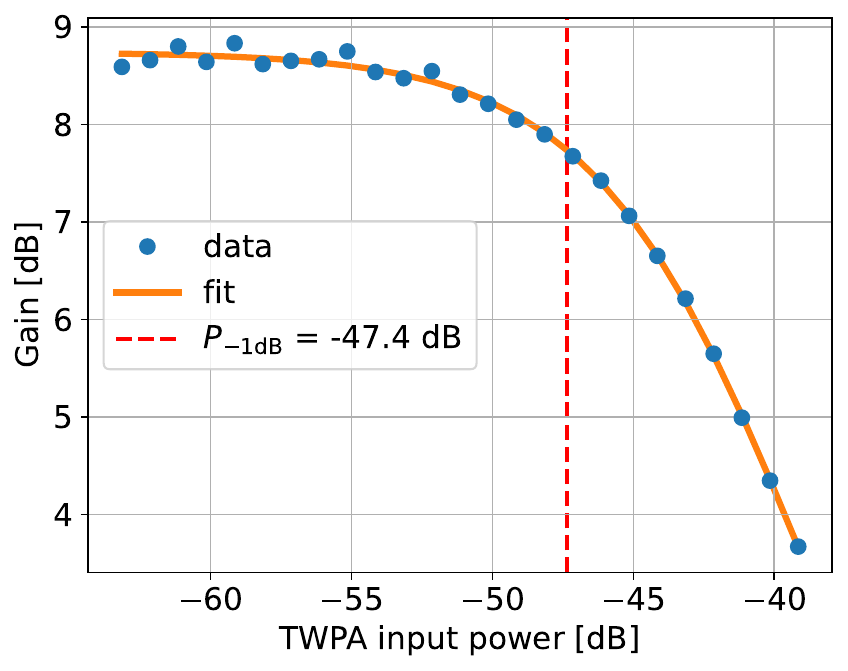}
    \caption{Measured gain as a function of the input signal power. From this plot, it is possible to extrapolate the $1\,$dB compression point $P_{-1\,\text{dB}}$, defined as the power where the gain has decreased of $1\,$dB in respect to its nominal value, due to saturation.}
    \label{fig:1db-compression-point}
\end{figure}

The final measurements focused on assessing the power handling capabilities of the device. While keeping the pump and dc bias parameters fixed, the mean gain was evaluated as a function of the VNA rf signal power injected into the TWPA. While the gain remains constant for low VNA powers, an increase in power causes the TWPA to saturate its output, leading to a gradual degradation of the gain.
It is common to extract the $1\,$dB compression point, which is identified as the input power necessary to deviate by $1$ dB from the maximum stable gain.
The measured mean gain as a function of the TWPA input power is shown in fig. \ref{fig:1db-compression-point}. A value of $P_{-1\text{dB}}=-47.4\,$dB was obtained considering the mean gain in the $3-5\,$GHz frequency range, which demonstrates the significant power handling capabilities of these devices with respect to other parametric amplifiers.

\section{Conclusions}\label{sec5}

This paper presented the latest developments of the DARTWARS project, regarding the advancements of KITWPAs for applications in quantum computing and sensing. We characterized one of the latest prototypes at cryogenic temperatures ($T\sim 60\,$mK) and measured a maximum mean gain of $9.11\,$dB in the frequency range from $3$ to $5\,$GHz.
The examined device exhibited a critical current of $I_c=(1.41\pm0.01)\,$mA and a scaling current of $I^*=(6.72\pm0.03)\,$mA, with its $1\,$dB compression point estimated at $P_{-1\text{dB}}=-47.4\,$dB.
The measured values are in reasonable agreement with the design goals of the prototype. Future phases of the project will involve noise performance characterization, producing additional short prototypes with an amplification bandwidth centered at higher frequencies, and eventually fabricating full-length devices with $20\,$dB of amplification.

\section*{Acknowledgement}
This work is supported by DARTWARS, a project funded by the Italian Institute of Nuclear Physics (INFN) within the Technological and Interdisciplinary Research Commission (CSN5), by European Union’s H2020-MSCA Grant Agreement No. 101027746, by the Italian National Centre for HPC Big Data and Quantum Computing (PNRR MUR project CN0000013-ICSC) and by the Italian National Quantum Science and Technology Institute (PNRR MUR project PE0000023-NQSTI). We acknowledge the support of the FBK cleanroom team for the fabrication. We also acknowledge useful discussions with Jiansong Gao, Michael Vissers, Jordan Wheeler, and Maxime Malnou. CB and SP acknowledge support from University of Salerno - Italy under the projects FRB19PAGAN, FRB20BARON and FRB22PAGAN.


\bibliography{sn-bibliography}

\end{document}